# In Silico Functional Profiling of Engineered Small Molecules: A Machine Learning Approach Leveraging PubChem Identifiers (CID_SID ML model)

Mariya L. Ivanova[1], *, Michael Nicholls[2], Nicola Russo[1], Gueorgui Mihaylov[3], Konstantin Nikolic[1]

Author affiliations

[1]School of Computing and Engineering, University of West London, London, UK

[2] The University of Law, London, UK

[3] Haleon, London, UK

*Corresponding author mariya.ivanova@uwl.ac.uk

## Abstract

The article introduces a concept for a time- and cost-effective methodological framework leveraging machine learning (ML) models for both early-stage drug development and clinical trial support. The rationale for this approach is the inherent scalability and speed enabled by using pre-calculated data embedded in existing PubChem identifiers (CID and SID), thereby eliminating the computationally intensive step of on-the-fly molecular descriptor generation. The approach was effectively demonstrated across four diverse bioassays: antagonists of the human D3 dopamine receptor, Rab9 promoter activators, small-molecule inhibitors of CHOP, and antagonists of the human M1 muscarinic receptor. A comparison, based on Matthews correlation coefficient (MCC), was conducted between the CID_SID ML model, the MORGAN2-based ML model, and the RDKit-transformed SMILES model for these four case studies, revealing that no method is universally superior in terms of performance. Furthermore, the **CID_SID model** averaged a rapid execution time of only **3.3 seconds**; the ML models relying on explicit structural descriptors, such as MORGAN2 and RDKit-transformed SMILES, demonstrated high computational costs, with processing times averaging 106.0 and 109.6 seconds, respectively. While negligible for a single ML model, these times would cause a significant difference in computational resource consumption when scaled across a framework involving over a million buildings. Moreover, the CID_SID ML model achieved strong average performance metrics: Accuracy of 83.52%±5.56%, Precision of 89.62%±4.91%, Recall of 75.65%±8.31%, F1-Score of 81.93%±6.76% and ROC of 83.53%±5.56%. So, the CID_SID ML model's decisive advantage in speed and reduced computational cost, combined with its reasonable metrics, makes it the more appropriate and robust building block for a scalable framework, particularly in scenarios that demand rapid triage and prioritisation of large compound libraries. This methodology prioritises the feasibility and operational efficiency of framework development without sacrificing scientific rigour.

**Key words**: Scikit-learn, Dopamine D3 receptor antagonists, Rab9 promoter activators, CHOP inhibitors, M1 muscarinic receptor antagonists.

## Introduction

The current process for bringing a new drug to market is costly and inefficient, with expenses exceeding USD 1 billion and timelines stretching over 10 years per successful candidate. (1) A major driver of this inefficiency is the high failure rate in clinical trials: 9 out of 10 drug candidates fail to gain FDA approval, frequently due to unexpected side effects, (2) so early screening for formulation liabilities is required. Discovering these liabilities early in development prevents significant financial and temporal investment in unviable candidates, thereby reducing the overall cost and duration of drug development.

To maximise clinical trial productivity, the protocol must be established as a Risk-Informed Protocol Foundation that integrates comprehensive, proactive measures to anticipate and mitigate the investigational compound's intrinsic risks. (3) This foundational approach necessitates early, deep analysis (often utilising in silico and in vitro modelling) of the compound's structure to predict potential undesired functionalities, such as off-target toxicity or adverse drug interactions. By mapping these risks prior to patient enrolment, the trial can be strategically designed with highly targeted inclusion/exclusion criteria, adaptive dose-escalation strategies, and strict, predefined safety-stopping rules. This preemptive strategy is crucial for preventing the significant time and financial waste associated with late-stage failures and for enabling efficient resource allocation, particularly through Risk-Based Monitoring that focuses on the highest-risk data points.

Computational approaches are widely acknowledged and firmly established across the field as methods that significantly accelerate and reduce the cost of drug discovery. In their review article, Ferreira and Carneiro analyse in detail how AI and ML are transforming the pharmaceutical pipeline and their role in drug discovery; synthesising the recent advancements (2019-2024), they comment on the virtual screening to clinical trial optimisation, covering RF-Score, BioBERT, AlphaFold, and advanced topics such as Explainable AI (XAI) and geometric deep learning. (4 -9) This positive and transformative trend is strongly supported by extensive evidence and numerous studies across various research domains, such as Structure-Based Drug Design (SBDD) and Ligand-Based Drug Design (LBDD); (10) the computational engineering of novel compounds` features, aiming to support the prediction of small biomolecule functionalities (Ivanova et al.); (11) using computational methods to find new uses for existing drugs (El-Atawneh and Goldblum, 2024); (12) applications involve using molecular dynamics and ML for designing drugs targeting the dopamine D3 receptor (Ferraro et al, 2020); (13) predicting receptor ligand selectivity; (14) applying structural similarity ML for M1 muscarinic receptor target identification. (15) However, while the execution of the ML models in the current study adhered to standard best practices (16), utilising ML library like the Scikit-learn (17) within a Jupyter notebook environment (18), the overall methodology is considered novel and unconventional because its central data-driven strategy focuses on employing features, descriptors encoded using the PubChem identifiers (compound identifier, CID  and substance identifier, SID), which are generally not used in conventional ML approaches for drug design. This reliance on an unconventional, identifier-based feature set, rather than purely structural or physicochemical properties, distinguishes the method from typical cheminformatics studies. Crucially, Generative AI (GenAI) was intentionally excluded. This decision was made to mitigate the risk of AI hallucination, thereby ensuring both the integrity and traceability of the results (19-20).

The current study proposes a methodology for development of building blocks for scalable predictive polypharmacology framework whose purpose is to enhance the efficiency of clinical trials significantly. The proposed approach identifies a compound's potential off-target interactions, which are the molecular origin of most toxic Adverse Events. By employing advanced in silico models, the framework would enable early elimination of high-risk candidates or the immediate establishment of stringent mitigation protocols. This critical foresight streamlines operations by enabling precise trial design, including targeted patient criteria, refined dose strategies, and the efficient deployment of Risk-Based Monitoring to focus resources solely on the most likely safety signals. (21) The result is a substantial reduction in costly late-stage failures and a more direct path to safe, effective market approval.

While a fundamental rule in ML is to avoid using sample Identifiers (IDs) during model training and evaluation due to their unsuitability for feature representation, the PubChem CIDs and SIDs present a notable exception in cheminformatics. (22-23) These are not arbitrary labels but sophisticated, structured keys that grant access to vast, curated chemical and biological data, making them practical for ML models. Specifically, the CID serves as the non-redundant identifier for a single, canonical chemical structure, established through a rigorous standardisation algorithm that resolves ambiguities such as tautomers and salts; this allows the CID to serve as a reliable anchor for all derived molecular properties, which are the true features used in ML. Furthermore, the SID is crucial as it links the canonical structure back to its specific experimental context and data source. By leveraging these identifiers, the ML model implicitly utilises PubChem's extensive data hierarchy and curation, enabling it to effectively group structurally and contextually similar compounds and contextualise activity data, a strategy proven successful in predictive modelling studies.

Since the study's primary objective was the computational cost, the comparison was not performed with the current state-of-the-art technique, Graph Neural Networks (GNNs), because this method is considered to be significantly more computationally demanding.(24) Instead, the comparison of the CID_SID ML model was specifically conducted against ML models based on Morgan2 fingerprints and RDKit-transformed SMILES, as these two methods are both widely accepted, computationally efficient benchmark standards within cheminformatics.(25, 26)  These two benchmark methods were chosen because their relatively low computational overhead allows for rapid model training and inference, making them appropriate baselines for evaluating the performance and resource footprint of the novel CID_SID ML model under investigation.

The methodology was demonstrated using four distinct PubChem bioassays. (27- 30) These assays, sourced via quantitative high-throughput screening (qHTS), provided the large volume of high-quality data critical for the development of robust ML models. (31, 32) Additionally, a bioassay focused on the aquasolubility of the compound was utilised as a tool for mitigating the severe imbalance of active and inactive compounds in the case study bioassays. (33)

### 1. PubChem AID 652054 bioassay focused on Dopamine Receptor D3 Antagonists (27)

This bioassay was primarily designed for the discovery of novel antagonists of the dopamine receptor D3, which is a therapeutic target for neuropsychiatric disorders such as addictions, schizophrenia, psychosis and L-DOPA-induced dyskinesias. (34, 35) The original dataset contained 364,367 rows with samples and 26 columns. Results were obtained using a luminometer reader with a 20-second exposure time. Compounds exhibiting an activity ≤-50 were considered active (9,117 samples), and those with an activity ≥-30 were considered inactive (339,862 samples). Compounds with inconclusive activity between these two values were not utilised. A comprehensive description is provided in the bioassay documentation. (27)

### 2. PubChem AID 485297 bioassay focused on Rab9 promoter activators (28)

This bioassay focused on identifying small chemical compounds that can modulate the expression of the endogenous protein Rab9, offering a potential new treatment modality for neurodegenerative lipidosis, including Nieman Pick Type C and Alzheimer's disease. (36) The original dataset contained 321,272 rows and 11 features. Tests were performed at compound concentrations of 2.3μM, 11.40μ, and 57.5μM, with results obtained by fitting the dose-response curve to the Hill equation. Compounds with an activity ≤ -50 were designated as **active**, and those with an activity ≥ -29 were designated as **inactive**. Refer to PubChem for comprehensive documentation. (**28**)

### 3. PubChem AID2732 bioassay focused on CHOP inhibitors (29)

The primary goal of this assay was to find small-molecule inhibitors for CHOP (DNA damage-inducible transcript 3). CHOP inhibition is hypothesised to prevent programmed Unfolded Protein Response (UPR) cell death, offering therapeutic application to conditions such as Alzheimer's disease, Parkinson's disease, haemophilia, lysosomal storage diseases, and diabetes. (37-41) The original dataset contained 219,165 rows and 10 features. Tests were performed at a 10microM compound concentration. Luminescence signal was measured using an Envision Multilabel plate reader and analysed via an algorithm. Compounds with an activity >70% were considered active (8,243 samples), and the remaining 210,921 compounds were labelled as inactive. Refer to PubChem for comprehensive documentation. (29)

**4. PubChem AID 588852 focused on M1 Muscarinic Receptor Antagonists (30)**

This bioassay (NIH, 2012) aimed to identify antagonists of the human M1 muscarinic receptor, which mediates Acetylcholine actions in the CNS and represents a target for treating conditions like cognition disorders, Alzheimer's disease, and schizophrenia. (42-45) The original dataset contained 359,484 rows and 9 features. Tests were performed at a 4microM compound concentration. Compounds were defined as active (4,590 samples) if their score exceeded a cutoff parameter, calculated as the sum of the average percent inhibition of the test compound wells and three times their standard deviation, with results normalized to 100%. Refer to PubChem for comprehensive documentation. (30)

**5.PubChem AID1996 bioassay focussed on aqueous solubility of small biomolecules** (33)

The significant class imbalance across the four primary bioassays was addressed through a filtering step, which was implemented using PubChem Bioassay 1996. This bioassay, which focuses on the water solubility of small molecules, served a crucial role as a filter, causing a reduction of inactive compounds across the case study bioassays. Although the AID 1996 dataset was found to contain 57,859 samples (40,860 soluble and 17,573 insoluble), the actual solubility labels were not used in the current study. The rationale for utilising the PubChem AID 1996 bioassay as a sift is further detailed in the section *Methodology*.

## Methodology

For each bioassay listed above, the PubChem CIDs, SIDs and activity results were extracted. As noted above, to mitigate the severe imbalance between active and inactive small biomolecules, each resulting dataset was merged with the water solubility dataset (PubChem AID 1996), retaining only the compounds common to both bioassays. The application of this strategy, which focused on the aqueous solubility of small biomolecules, introduced a deliberate selection bias. This bias was considered necessary so that the modelling efforts could be concentrated on the chemical space most conducive to drug discovery success, with compounds possessing favourable and reliable physicochemical properties being particularly targeted. As noted above, while substantial data were offered by the AID 1996 dataset (57,859 total rows: 40,860 soluble, 17,573 insoluble), the explicit solubility information was ignored. Only the CIDs were extracted from AID 1996, and only the samples shared with the considered target PubChem bioassays (PubChem AID 652054, AID485297, AID2732, and AID588852) were retained. This step was taken specifically to ensure that the number of inactive compounds in the targeted bioassays was reduced. Additionally, inactive compounds were reduced by extracting every n-th record, where the value of n was individually determined based on the number of active small molecules in the relevant bioassay. The final datasets were formed by a combination of the set with the reduced inactive compounds and all active compounds from the targeted bioassays. Through this combined approach, the simultaneous mitigation of data imbalance and the management of the selection bias were achieved.

To prepare the datasets for ML, an equal number of targets (Class 1 and Class 0), corresponding to 10% of the entire final dataset for each class, were extracted and concatenated. This resulted in a 20% test dataset with an equal number of targets, ensuring that a reliable evaluation of the models will be achieved. Using an equal number of samples from each class in the test set (a balanced, randomly selected test set) ensures a fair and reliable evaluation of the ML model's ability to generalise across all outcomes. This strategy is critical because it directly addresses the accuracy paradox, in which an imbalanced test set can yield a deceptively high accuracy score from a model that predicts the majority class, thereby masking poor performance on the minority class. By enforcing equal class proportions, the model is compelled to correctly identify instances from every class, ensuring the overall accuracy is a meaningful metric that reflects equal emphasis and penalty for misclassification across all outcomes. Furthermore, a balanced test set facilitates straightforward, reliable interpretation of standard performance metrics. The remaining compounds were used to create the train set. Initially, the datapoints of these sets were scaled, and then, along with the target part of the train sets, were balanced. The balancing was performed using either the Synthetic Minority Oversampling Technique (SMOTE) or the Random Oversampling (ROS). (46, 47)

ML was performed with algorithms provided by scikit-learn, which included: Decision Tree Classifier (DTC), Random Forest Classifier (RFC), Support Vector Classifier (SVC), Gradient Boosting Classifier (GBC) and XGBoosting Classifier (XGB). (48, 49) Cross-validation was used to estimate how well a model generalises to unseen data. (50) Subsequently, by comparing the train and test accuracy, the best-performing model was scrutinised for overfitting, identifying the moment for early stopping in order to increase the generalisation ability. (51) A deviation between the test and train accuracy greater than 5% was accepted as an indicator of when overfitting begins. The identification moment for early stopping was used during further hyperparameter tuning to improve the performance of the ML models.

Two distinct approaches for hyperparameter tuning were explored: the first involved grid search, and the second was real-time hyperparameter tuning using the Optuna API based on Bayesian optimisation. (52, 53)The appropriate hyperparameters were chosen according to the type of ML model. Subsequently, after the hyperparameters tuning, the models with the newly tuned hyperparameter were scrutinised as well for overfitting.

Since the ML models were classification, the ML metrics used for their evaluations were Accuracy (percentage of correct predictions divided by the total number of predictions); metrics used for the evaluation of the ML models in the study were: (54)

- Accuracy: The percentage of correct predictions divided by the total number of predictions.
- Precision: The accuracy of positive predictions made by a model.
- Recall: All actual positive instances predicted within a dataset.
- F1-score: Harmonic mean combining Precision and Recall.
- ROC: The ability of a binary classifier system when its discrimination threshold is varied.

The structure of the methodology is outlined in Figure 1.

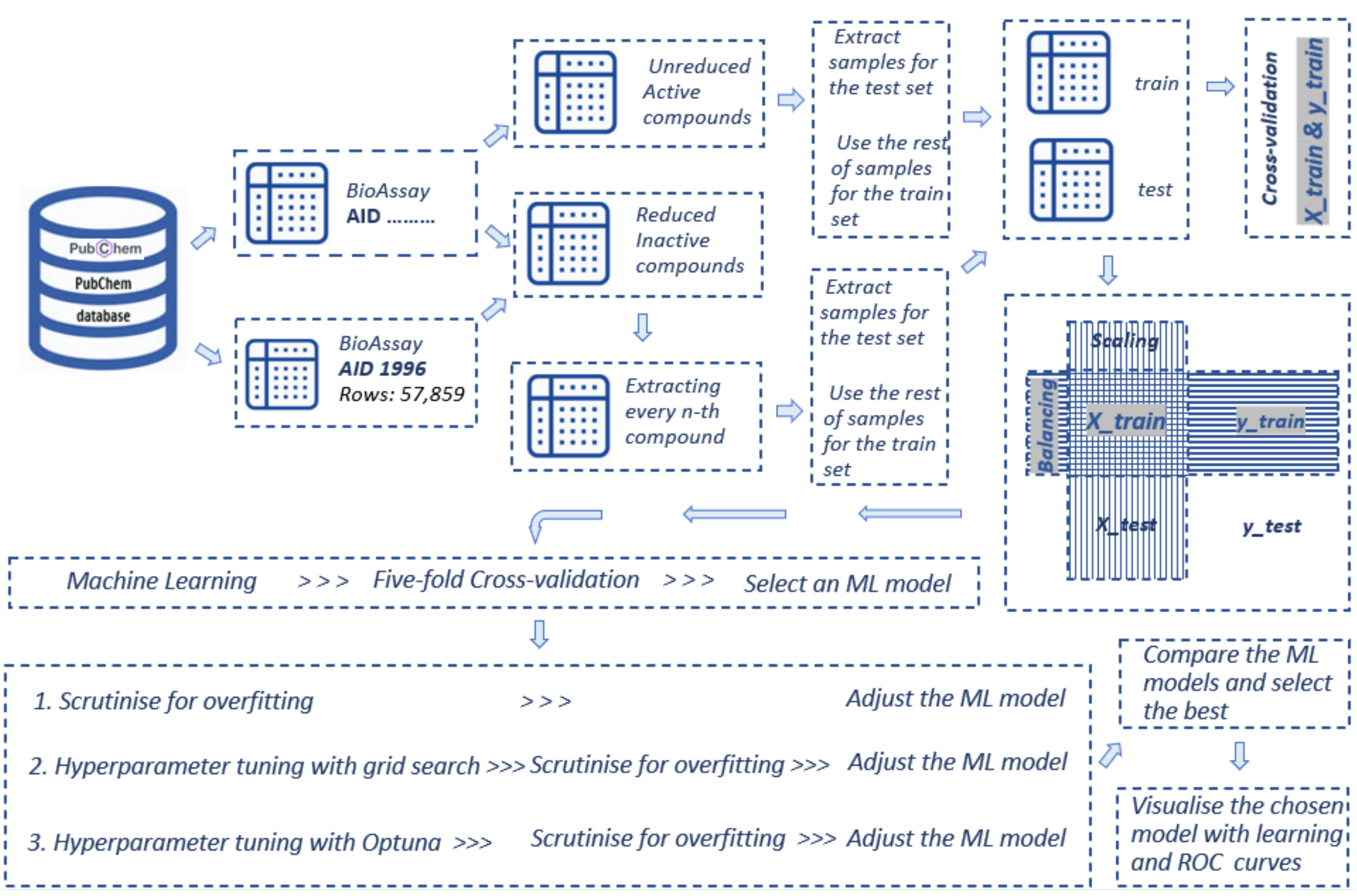

Figure 1. Methodology for development of a CID-SID ML model

To validate the final models and ensure a thorough assessment of their performance and reliability, several rigorous evaluation steps were undertaken in the methodology. The confusion matrix and classification report were plotted to obtain a complete breakdown of class-specific performance metrics, including Precision, Recall, and F_1-score, which were considered essential for a comprehensive evaluation beyond simple Accuracy. Furthermore, the learning curve was visualised so that the model's behaviour regarding training data size and potential overfitting could be diagnosed, while the ROC curve was plotted to evaluate the discriminatory power of the classifier across various thresholds. Lastly, LIME (Local Interpretable Model-agnostic Explanations) analysis was used to provide local interpretability for specific predictions (Ribeiro et al., 2016). (55) The model's robustness was explored by introducing Gaussian noise and missing values to ensure its stability when confronted with corrupted or incomplete real-world data.

The scientific context and performance proof were established through comparisons of the CID_SID ML model against two industry-standard baseline methodologies: MORGAN2 fingerprints and an ML model based on RDKit-transformed SMILES. (25, 26) The RDKit representations were essential because they provide a proven, accurate, and highly computationally efficient means of translating complex molecular information into a fixed numerical format that is readily processed by ML and Quantitative Structure-Activity Relationship (QSAR) algorithms. The comparison was conducted by evaluating each ML model based on two key metrics: the Matthews Correlation Coefficient (MCC) and the computational time required for performance, including feature transformation, scaling, balancing, training, predicting and evaluation. (56)

## Results and discussion

1. **Results of the CID_SID ML model based on PubChem Bioassay AID652054, focusing on human Dopamine D3 receptor antagonists.**

The final dataset was constructed by merging the PubChem AID 652054 Bioassay (364,367 samples) with the PubChem AID 1996 Bioassay (57,859 samples), resulting in a dataset of 55,780 samples in total, 54,951 of which were inactive small biomolecules and 829 active. (27, 33) To continue reducing the inactive compound population, a systematic sampling approach was applied; every fourth sample from the population was extracted, creating a reduced set of 13,738 inactive samples. These were then integrated with all 9,117 active compounds to generate a final, consolidated dataset of 22,855 compounds for further use in the study. (57)

To ensure equal target representation and maximise model robustness, the test set was created by removing 10% of the active and inactive compounds, respectively. Specifically, 1,910 samples were randomly set aside from the inactive compounds after shuffling, and the same was done for the active compounds. These two classes were then integrated to form a balanced test set of 3,820 compounds, representing 20% of the initial consolidated dataset and featuring an equal proportion of targets. The remaining 19,035 compounds (22,855 total - 3,820 test) were designated for the training set. This training set was further processed by applying both scaling and balancing techniques (as illustrated in Figure 1). The imbalance in the training data was addressed using a RandomOverSampler, which effectively increased the number of compounds from 19,035 to 23,656 compounds, consequently ensuring the model was trained on a fully balanced dataset.

The metrics of the ML models listed in the Methodology section were compared, revealing that the XGBC was the most effective estimator, achieving the highest test set metrics: 85.6% Accuracy and 85.6% ROC, followed by RFC with 85.2 % Accuracy and 85.1% ROC (Table 1).

Table 1. Metrics across the ML models based on the dopamine D3 receptor antagonist data

| Algorithm | Accuracy | Precision | Recall | F1 | ROC |
|---|---|---|---|---|---|
| **XGBoost** | 0.856 | 0.901 | 0.799 | 0.847 | 0.856 |
| **RandomForest** | 0.852 | 0.905 | 0.787 | 0.842 | 0.852 |
| **GradientBoost** | 0.848 | 0.902 | 0.781 | 0.837 | 0.848 |
| **Decision** | 0.832 | 0.861 | 0.791 | 0.825 | 0.832 |
| **K-nearest** | 0.830 | 0.843 | 0.812 | 0.827 | 0.830 |
| **SVM** | 0.808 | 0.904 | 0.688 | 0.781 | 0.808 |

Further analysis using five-fold cross-validation confirmed XGBC's robustness, yielding an average Accuracy of 86.45%±0.59, Precision of 84.60%±0.61, Recall of 78.49%±1.58, F1-score of 81.42%±0.95 and ROC AUC of 91%±0.62. The plot (Figure 2.a) illustrates the impact of the max_depth hyperparameter on model performance, specifically highlighting the onset of overfitting relative to a 5% (0.05) deviation threshold between the Training and Test Accuracy. At max_depth = 6, the deviation is exactly 0.906 - 0.856 = 0.050, placing the model right at the acceptable limit for complexity. However, increasing the depth to max_depth = 7 causes the deviation to grow to 0.916 - 0.850 = 0.066, which clearly exceeds the 5% threshold. This sharp increase in the generalisation gap between depth 6 and 7 confirms that the model starts to significantly overfit the training data past max_depth=6, sacrificing meaningful performance on unseen data for marginal gains in training accuracy. Therefore, max_depth=6 represents the optimal balance point before excessive complexity compromises robustness. **The final XGBC, hyperparameter-tuned and checked for overfitting, achieved an Accuracy of 85.6%, a Precision of 89.4%, a Recall of 80.8%, an F1-score of 84.9% and an ROC of 85.6%.**

The provided learning curve (Figure 2.b) clearly demonstrates significant overfitting when the training dataset size is small, characterised by a large gap between the near-perfect Training Score (1.0) and the low Cross-Validation Score (0.5) observed early on. As the number of training examples increases, the model's generalisation ability improves sharply, evidenced by the rising cross-validation score. However, past approximately 15,000 examples, both the training score (which drops to around 0.9) and the cross-validation score (which stabilises around 0.86) begin to plateau. This final state suggests that while increasing the data size was beneficial, the model still suffers from high variance, as indicated by the remaining separation between the two curves. To achieve better generalisation performance beyond the current 86%, further efforts should focus on reducing model complexity or improving feature quality rather than solely relying on additional data.

The ROC curve for the XGBC demonstrates excellent discriminatory power, evidenced by an Area Under the Curve (AUC) of 0.91, indicating a 91% probability that the model correctly ranks a random positive instance higher than a random negative instance. The curve's steep ascent toward the upper-left corner confirms the model's robustness, showing that a high True Positive Rate (TPR) can be achieved while maintaining a very low False Positive Rate (FPR). For instance, a TPR of approximately 0.8 is achieved with an FPR of less than 0.1, meaning that the model is substantially better than a random guess (AUC = 0.5) and is highly effective at distinguishing between the two classes.

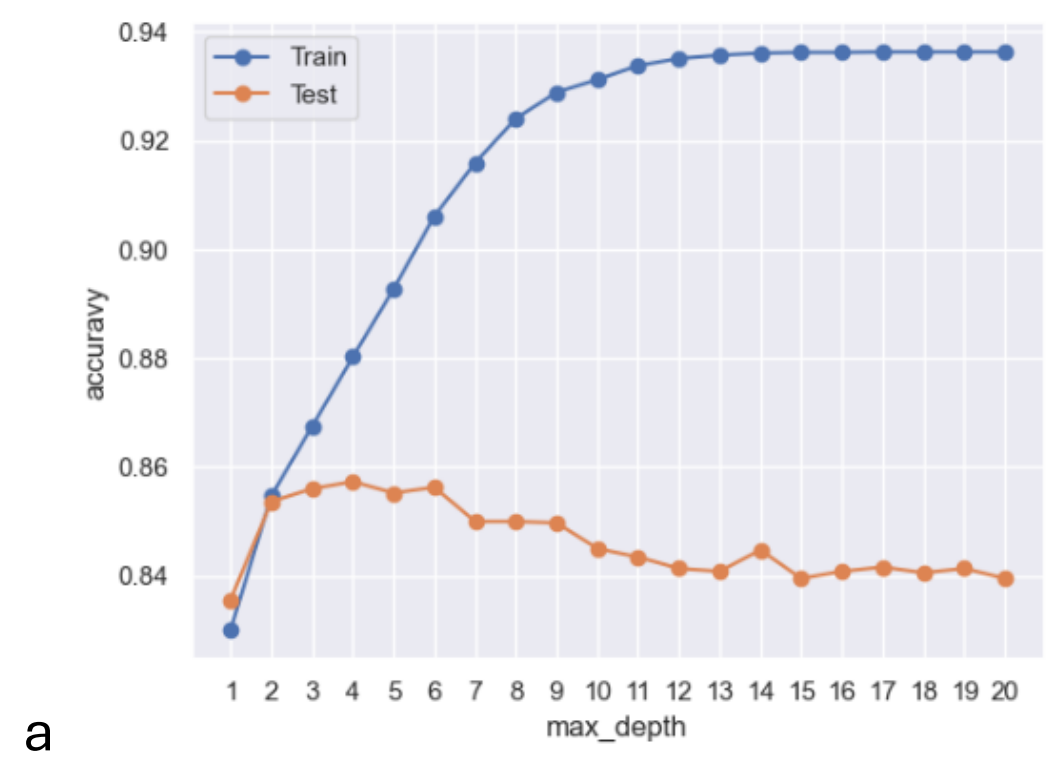

a

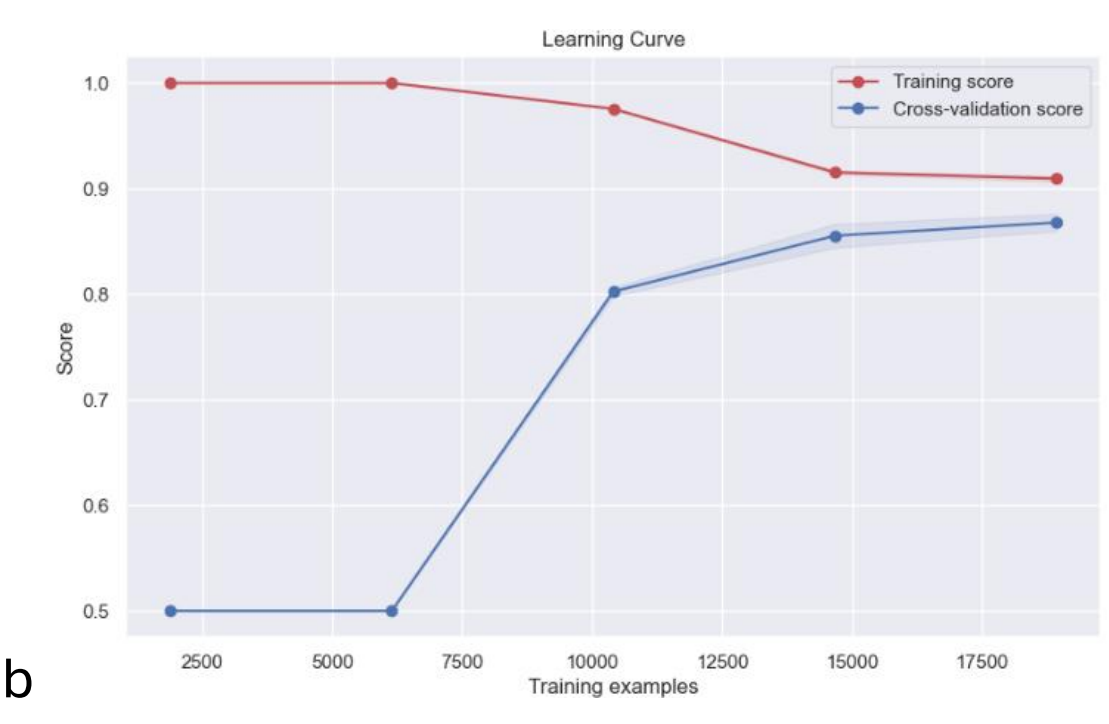

b

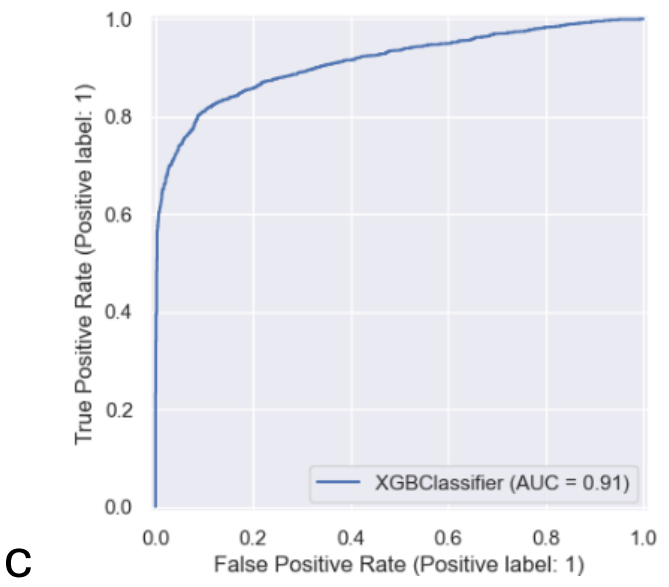

c

Figure 2. CID_SID ML model based on PubChem Bioassay AID652054, focusing on human Dopamine D3 receptor antagonists.

a) Tracing the deviation between train and test accuracy
b) Learning curve
c) AUC

This classification report demonstrates that the model is highly effective and well-balanced, achieving a strong overall Accuracy of 0.86 (86%) on a perfectly balanced dataset where both the Active (target 1) and Inactive (target 0) classes have equal support (1910 samples). The high F1-scores (Active=0.86, Inactive=0.85) confirm the model's excellent discriminatory power for both classes. While the performance is balanced, there is a slight, strategic trade-off: the model prioritises Recall on the Active class (0.90), indicating a notable ability to correctly identify true positive instances (minimising false negatives), and prioritises Precision on the Inactive class (0.89), demonstrating that its negative predictions are highly reliable. This consistent performance is reflected in the high macro and weighted averages (both 0.86), which suggests that the model is robust and well-generalised for this binary classification task.

Table 2. Classification report for XGBC based on PubChem Bioassay AID652054, focusing on human Dopamine D3 receptor antagonists.

| | precision | recall | F1-score | support |
|---|---|---|---|---|
| Active (target 1) | 0.82 | 0.90 | 0.86 | 1910 |
| Inactive (target 0) | 0.89 | 0.81 | 0.85 | 1910 |
| | | | | |
| accuracy | | | 0.86 | 3820 |
| macro avg | 0.86 | 0.86 | 0.86 | 3820 |
| Weighted avg | 0.86 | 0.86 | 0.86 | 3820 |

The XGBC demonstrated varied resilience across the two data corruption scenarios, resulting in an Overall Average Robustness Accuracy of 0.6666 across all ten tests. The model showed poor robustness to Gaussian Noise, with its accuracy sharply declining from 0.6830 at the 0.1 noise level to plateauing near 0.57 at the 0.6 level. This steep performance drop indicates that the model is unstable to random feature perturbations, suggesting that its decision boundaries rely heavily on precise input values.

Conversely, the model exhibited significantly better resilience to Missing Values Handled by Imputation, maintaining an accuracy of 0.8236 at a 0.1 missing rate and degrading gradually to 0.6992 at a 0.4 rate. This suggests that the tree-based structure and the imputation strategy successfully mitigate the impact of missing entries, making the model more robust to data sparseness than to random corruption of existing features.

The LIME result successfully validates the XGBC's correct prediction of True Label 0 (the Negative Class) for the specific instance index 1, with overall model accuracy 0.8552. The local explanation identifies that the features 'SID > 0.37' and 'CID > -0.08' were the top two influences driving the prediction toward class 0, contributing weights of approximately 0.147 and 0.105, respectively. These significant, positive contributions show that in the localised region of the feature space surrounding this specific sample, the model's decision was strongly dictated by these two clear-cut feature conditions.

The CID_SID ML mode emerged as the clearly superior performer with the highest MCC of 0.7159, demonstrating the best overall balance in classification by achieving the highest TP (1543) and the lowest FP (182) and FN (367) relative to its sample size (Figure 3). Among the two models tested on the larger dataset, the RDKit SMILES ML based model proved to be the stronger contender (MCC: 0.6528) due to its superior ability to correctly identify the negative class (highest TN: 2432) and its lower FN (439) compared to MORGEN2 based ML model (MCC: 0.5833), which suffered from the lowest TP(1272) and highest FN (551), indicating a significant struggle in accurately predicting the positive class. In summary, the CID-SID model provides the best predictive balance; the RDKit model offers the strongest performance among the descriptor-based alternatives tested on the larger data set.

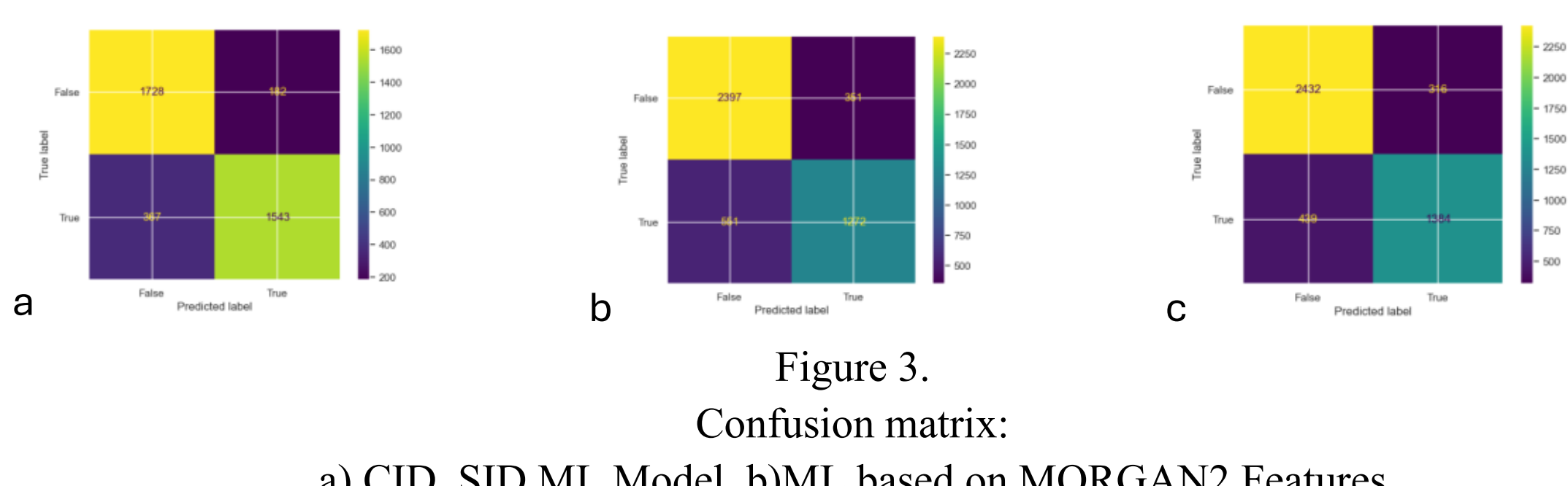

Figure 3.
Confusion matrix:
a) CID_SID ML Model, b)ML based on MORGAN2 Features,
c)ML based on RDKit transformed SMILES

2. **Results of the CID_SID ML model based on PubChem bioassay AID485287, focusing on promoter Rab9 activators**

The final dataset was constructed by merging the PubChem AID 485287 Bioassay (321,292 samples) with the PubChem AID 1996 Bioassay (57,859 samples), resulting in a dataset of 47,918 samples in total, 45,402 of which were inactive small biomolecules and 2,516 active. (28, 33) To continue reducing the inactive compound population, a systematic sampling approach was applied; every fourth sample from the population was extracted, creating a reduced set of 11,351 inactive samples. These were then integrated with all 9,138 active compounds to generate a final, consolidated dataset of 20,489 compounds for further use in the study. (58)

To ensure equal target representation and maximise model robustness, the test set was created by removing 10% of the active and inactive compounds, respectively. Specifically, 2,050 samples were randomly set aside from the inactive compounds after shuffling, and the same was done for the active compounds. These two classes were then integrated to form a balanced test set of 4,100 compounds, representing 20% of the initial consolidated dataset and featuring an equal proportion of targets. The remaining 16,389 compounds (20,489 - 4,100 test) were designated for the training set. This training set was further processed by applying both scaling and balancing techniques (as illustrated in Figure 1). The imbalance in the training data was addressed using a RandomOverSapmpler, which effectively increased the number of compounds from 16,389 to 18,602 compounds, consequently ensuring the model was trained on a fully balanced dataset.

The metrics of the ML models listed in the Methodology section were compared, revealing that the GBC was the most effective estimator, achieving the highest test set metrics: 75.4% Accuracy and 75.4% ROC, followed by GBC with 75% Accuracy and 75% ROC (Table 3).

Table 3. Metrics across the ML models based on PubChem bioassay AID485287, focusing on promoter Rab9 activators

| Algorithm | Accuracy | Precision | Recall | F1 | ROC |
|---|---|---|---|---|---|
| GradientBoost | 0.754 | 0.827 | 0.641 | 0.722 | 0.754 |
| XGBoost | 0.750 | 0.869 | 0.588 | 0.701 | 0.750 |
| RandomForest | 0.747 | 0.792 | 0.670 | 0.726 | 0.747 |
| K-nearest | 0.726 | 0.740 | 0.697 | 0.718 | 0.726 |
| SVM | 0.722 | 0.972 | 0.457 | 0.622 | 0.722 |
| Decision | 0.714 | 0.731 | 0.677 | 0.703 | 0.714 |

Further analysis using five-fold cross-validation confirmed XGBC's robustness, yielding an average Accuracy of 77.13%±0.83, Precision of 85.09%±1.98, Recall of 57.17%±1.06, F1-score of 68.38%±1.07 and ROC AUC of 81.16%±0.59. The plot (Figure 2, a) illustrates the impact of the max_depth hyperparameter on model performance, specifically highlighting the onset of overfitting relative to a 5% (0.05) deviation threshold between the Training and Test Accuracy. At max_depth = 2, the deviation is exactly 0.781 - 0.746 = 0.035, placing the model right at the acceptable limit for complexity. However, increasing the depth to max_depth = 3 causes the deviation to grow to 0.815 - 0.750 = 0.065, which clearly exceeds the 5% threshold. This sharp increase in the generalisation gap between depth 6 and 7 confirms that the model starts to significantly overfit the training data past max_depth=6, sacrificing meaningful performance on unseen data for marginal gains in training accuracy. Therefore, max_depth=2 represents the optimal balance point before excessive complexity compromises robustness. **The final GBC, hyperparameter-tuned and checked for overfitting, achieved an Accuracy of 75%, a Precision of 84.3%, a Recall of 61.14%, an F1-score of 71% and an ROC of 75%.**

The learning curve(Figure 2.b) for the GBC provides a precise indication of its bias-variance characteristics. Initially, at low sample sizes (up to ~6,000), the model displays severe overfitting, evidenced by the considerable gap between the near-perfect Training Score ( ~1.0) and the very low Cross-validation Score (~0.50). As the training data size approaches the maximum (~20,000 samples), the curves swiftly converge; the Training Score decreases while the Cross-validation Score rises sharply. This convergence indicates that the model's variance substantially lower, forcing it to generalise effectively. Ultimately, both scores plateau around 0.83, suggesting the GBC has achieved an optimal fit for the current dataset size and complexity. Further improvement in generalisation performance is unlikely through the sole addition of more training examples.

Figure 3.c, which displays the DOC curve for the GBC, indicates strong discriminatory performance. The key finding is the AUC value of 0.82. Since the AUC ranges from 0.5 (random chance) to 1.0 (perfect classification), an AUC of 0.82 signifies that the model is highly effective at distinguishing between the positive and negative classes. Specifically, there is an 82% chance that the model will correctly rank a randomly selected positive sample higher than a randomly picked negative sample. The curve's steep ascent and proximity to the top-left corner visually confirm this result, demonstrating that the model can achieve a high TP rate while maintaining a low FP rate across various classification thresholds.

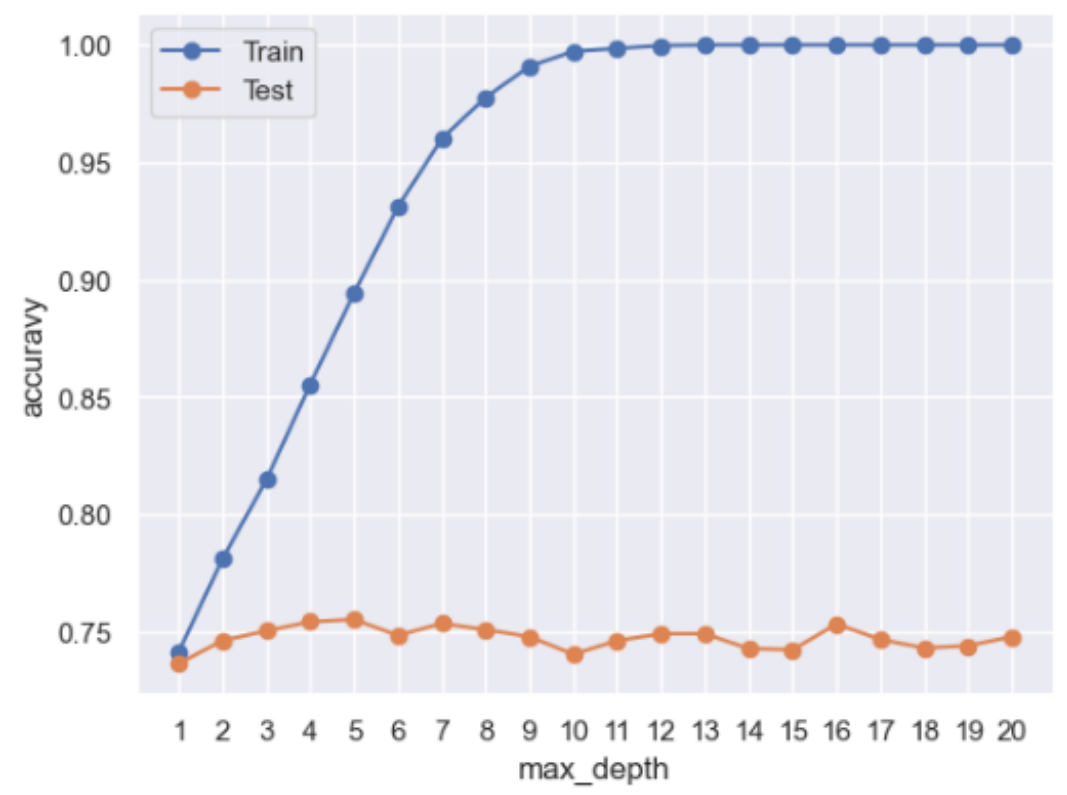

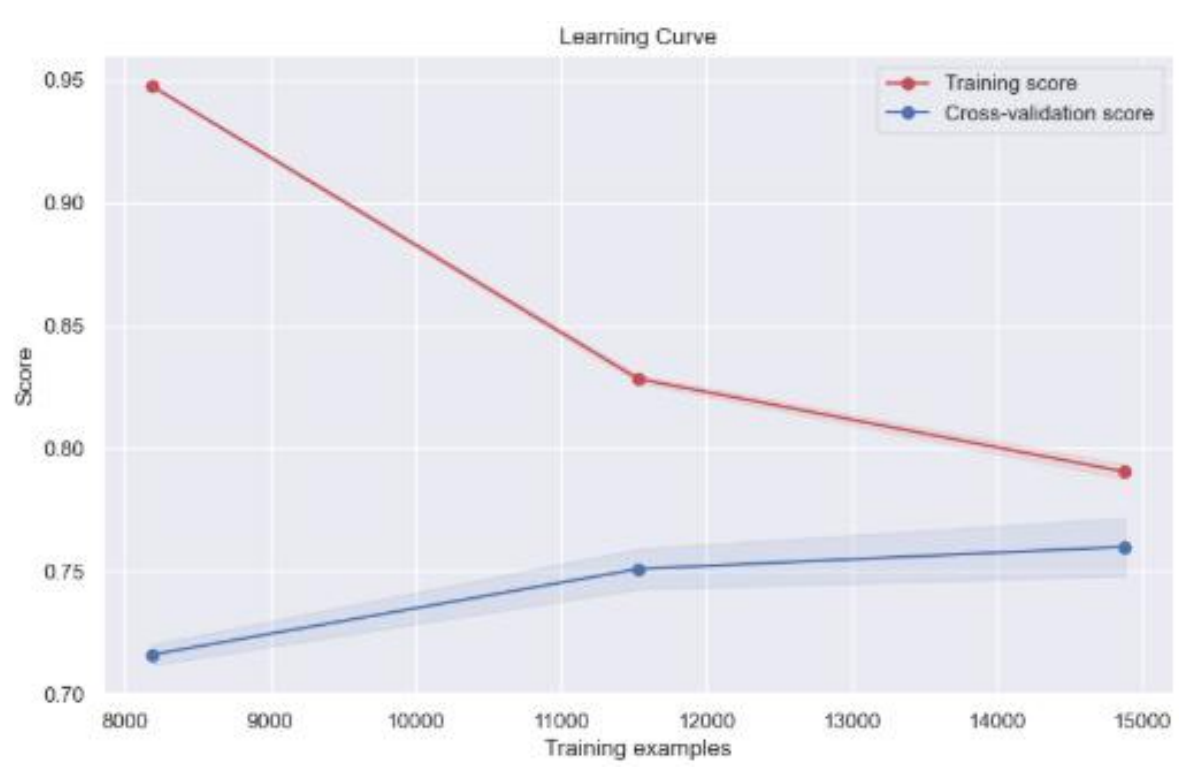

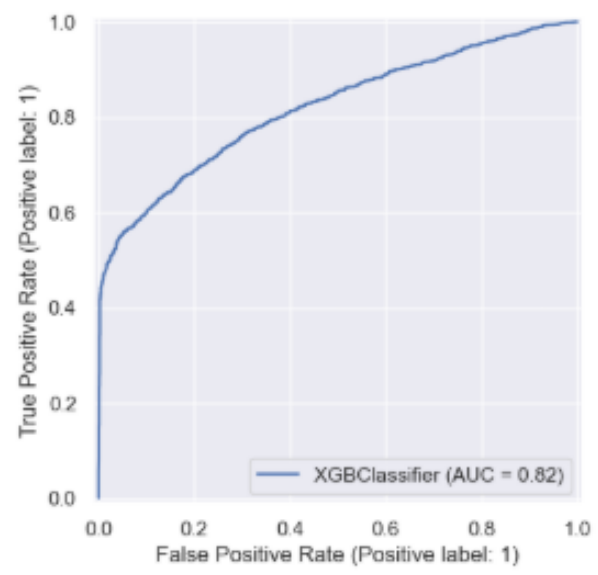

Figure 5. CID_SID ML model based on PubChem bioassay AID485287, focusing on promoter Rab9 activators

a) Tracing the deviation between train and test accuracy
b) Learning curve
c) AUC

The classification report for the GBC, leveraging CID and SID data, indicates a balanced overall model performance, achieving a global **accuracy of 75%** across the 4,100 evaluated samples. Notably, the dataset appears perfectly balanced, with equal support for both the Active (target 1) and Inactive (target 0) classes (2,050 samples each). The model exhibits a clear bias towards the Active (target 1) class, achieving a very high Recall of 88%. This indicates strong sensitivity, meaning it accurately identifies nearly nine out of ten actual active compounds, which is crucial for reducing false negatives in screening applications. However, this sensitivity comes at the cost of precision (70%) for the Active class, suggesting that 30% of the compounds predicted as active are actually inactive (false positives). Conversely, the model shows stronger Precision (84%) for the Inactive (target 0) class, but its Recall is significantly lower at 62%. This implies the model is more reliable for predicting inactivity, but it misses a larger proportion of actual inactive compounds. The resulting macro and weighted average F1-scores of 0.74 confirm the model's solid, though not exceptional, performance, highlighting the trade-off between the high sensitivity for the Active class and the higher specificity for the Inactive class.

Table 4. Classification report for GBC based on PubChem bioassay AID485287, focusing on promoter Rab9 activators

| | precision | recall | F1-score | support |
|---|---|---|---|---|
| Active (target 1) | 0.70 | 0.88 | 0.78 | 2050 |
| Inactive (target 0) | 0.84 | 0.62 | 0.71 | 2050 |
| | | | | |
| accuracy | | | 0.75 | 4100 |
| macro avg | 0.77 | 0.75 | 0.74 | 4100 |
| Weighted avg | 0.77 | 0.75 | 0.74 | 4100 |

The robustness analysis comparing the model's performance on clean data (Baseline Accuracy: 0.7534) against corrupted data reveals a significant drop in predictive reliability, confirming the model is not robust to real-world data imperfections. The model shows the greatest sensitivity to Gaussian Noise, where accuracy plummets by ~17.5% to 0.6217 at the lowest noise level, indicating poor tolerance for random feature perturbations. While performance under Missing Value Imputation is better, it still sees a significant drop, with accuracy falling to 0.7049 when 10% of data is missing and continuing to decline. Consequently, the Overall Average Robustness Accuracy is 0.6075, a substantial 14.59% reduction from the baseline. This indicates that despite strong performance on curated data, the model's operational stability is compromised in environments with typical data uncertainty, necessitating better data cleaning or the adoption of more robust feature engineering techniques.

The LIME analysis for the GBC focuses on a successful classification (TN) for Instance 0, where both the True Label and Model Prediction were 0 (Inactive). The model's decision was overwhelmingly driven by the feature $-0.38 < \text{SID} \leq 1.08$, which contributed a strong negative weight of -0.3795. This confirms that the GBC correctly learned to associate intermediate values of the SID feature with the Inactive class in this local region of the feature space. The other feature, $\text{CID} \leq -0.34$, played only a minor and counteractive role, providing a small positive influence (+0.0990), underscoring the dominance of the SID feature in this specific classification instance.

The comparison of the three binary classification models based on their MCC reveals that the RDKit-transformed SMILES model is marginally the optimal classifier with an MCC of 0.6392. This model achieves the highest number of TP (1533) while maintaining a competitive FN count (295) (Figure 5). Achieving an MCC of 0.6377, the MORGAN2

Features model closely trails the leader. While it excels in specificity with the highest count of TN (1852), its high FP count (418) represents a significant source of error. In contrast, the CID_SID ML Model is the weakest performer (MCC ~0.5143), suffering from a critically high number of FN (787), which indicates a severe inability to correctly identify the positive class despite a good FP rate, the RDKit and MORGAN2 models demonstrate better overall balance, with the RDKit approach providing the best holistic classification performance.

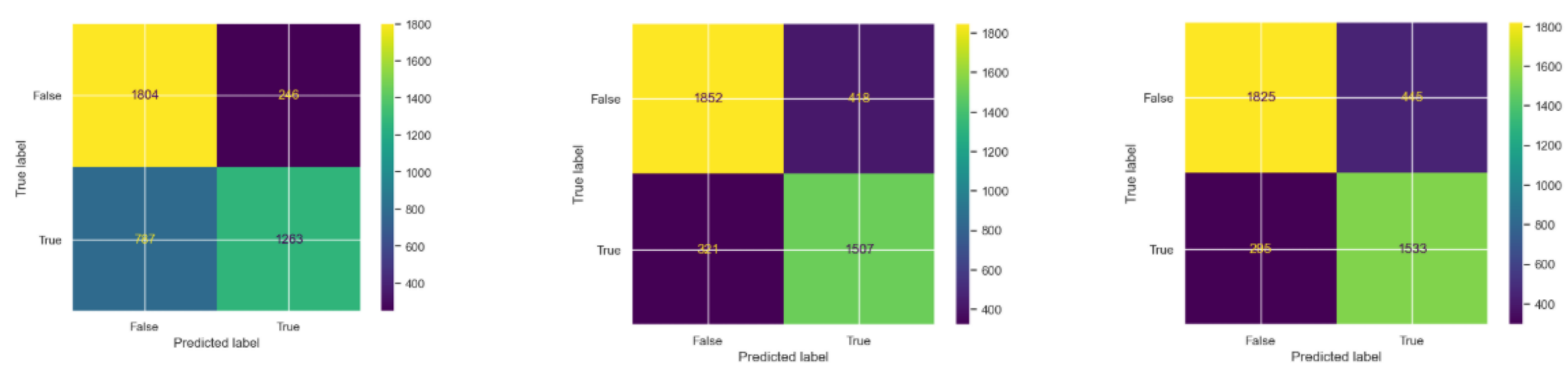

Figure 5.
Confusion matrix:
a) CID_SID ML Model, b)ML based on MORGAN2 Features,
c)ML based on RDKit transformed SMILES

3. **Results of the CID_SID ML model based on PubChem bioassay AID2735, focusing on CHOP inhibitors**

The final dataset was constructed by merging the PubChem AID 2732 Bioassay (219,165 samples) with the PubChem AID 1996 Bioassay (57,859 samples), resulting in a dataset of 24,480 samples in total, 24,188 of which were inactive small biomolecules and 292 active. (29, 33) To continue reducing the inactive compound population, a systematic sampling approach was applied; every fourth sample from the population was extracted, creating a reduced set of 11,351 inactive samples. These were then integrated with all 8,243 active compounds to generate a final, consolidated dataset of 32,431 compounds for further use in the study. (59)

To ensure equal target representation and maximise model robustness, the test set was created by removing 10% of the active and inactive compounds, respectively. Specifically, 3,250 samples were randomly set aside from the inactive compounds after shuffling, and the same was done for the active compounds. These two classes were then integrated to form a balanced test set of 6,500 compounds, representing 20% of the initial consolidated dataset and featuring an equal proportion of targets. The remaining 25,931compounds (32,431 – 6,500 test) were designated for the training set. This training set was further processed by applying both scaling and balancing techniques (as illustrated in Figure 1). The imbalance in the training data was addressed using a RandomOverSapmpler, which effectively increased the number of compounds from 25,931 to 41,876 compounds, consequently ensuring the model was trained on a fully balanced dataset.

The metrics of the ML models listed in the Methodology section were compared, revealing that the XGBC was the most effective estimator, achieving the highest test set metrics: 75.4% Accuracy and 75.4% ROC, followed by GBC with 75% Accuracy and 75% ROC (Table 5).

Table 5 . Metrics across the ML model based on PubChem bioassay AID2735, focusing on CHOP inhibitors

| Algorithm | Accuracy | Precision | Recall | F1 | ROC |
|---|---|---|---|---|---|
| GradientBoost | 0.902 | 0.983 | 0.818 | 0.893 | 0.902 |
| XGBoost | 0.897 | 0.961 | 0.827 | 0.889 | 0.897 |
| RandomForest | 0.893 | 0.966 | 0.814 | 0.884 | 0.893 |
| SVC | 0.884 | 0.980 | 0.784 | 0.871 | 0.884 |
| Decision | 0.883 | 0.955 | 0.804 | 0.873 | 0.883 |
| K-nearest | 0.876 | 0.909 | 0.837 | 0.871 | 0.876 |

Further analysis using five-fold cross-validation confirmed XGBC's robustness, yielding an average Accuracy of 86.45%±0.59, Precision of 84.60%±0.61, Recall of 78.49%±1.58, F1-score of 81.42%±0.95 and ROC AUC of 91%±0.62. The plot (Figure 2, a) illustrates the impact of the max_depth hyperparameter on model performance, specifically highlighting the onset of overfitting relative to a 5% (0.05) deviation threshold between the Training and Test Accuracy. At max_depth = 6, the deviation is exactly 0.906 - 0.856 = 0.050, placing the model right at the acceptable limit for complexity. However, increasing the depth to max_depth = 7 causes the deviation to grow to 0.916 - 0.850 = 0.066, which clearly exceeds the 5% threshold. This sharp increase in the generalisation gap between depth 6 and 7 confirms that the model starts to significantly overfit the training data past max_depth=6, sacrificing meaningful performance on unseen data for marginal gains in training accuracy. Therefore, max_depth=6 represents the optimal balance point before excessive complexity compromises robustness. **The final XGBC, hyperparameter-tuned and checked for overfitting, achieved an Accuracy of 85.6%, a Precision of 89.4%, a Recall of 80.8%, an F1-score of 84.9% and an ROC of 85.6%.**

Figure 6.b. displays the Learning Curve for the GBC, which exhibits a general pattern of low bias but benefits significantly from increased data to reduce variance. Over the range of training examples from 18,000 to 34,000, both the Training Score (Red Line) and the Cross-validation Score (Blue Line) maintain very high values (above 0.89), confirming the model's inherent complexity and predictive capability. A persistent, though small, gap exists between the scores, indicating slight overfitting (variance), but this gap systematically narrows as the training set size increases. Critically, the Cross-validation

Score is still visibly rising at the maximum data point ( ~34,000 with a score of ~0.92), suggesting that the model has not yet reached its performance ceiling. The primary conclusion is that adding more training data would likely lead to further improvements in the generalization score and fully mitigate the remaining variance.

Figure 6.c, which displays the ROC curve for the GBC, indicates exceptional discriminatory performance as summarized by the AUC value of 0.95. Since the AUC ranges from 0.5 (random guessing) to 1.0 (perfect classification), this score signifies that the model has a 95% chance of correctly ranking a randomly selected positive instance higher than a randomly chosen negative instance. The visual representation confirms this strength, as the curve rises very steeply and remains extremely close to the top-left corner (the ideal point of high TP Rate and low FP Rate), demonstrating the model's ability to maintain high sensitivity while achieving high specificity across all practical classification thresholds.

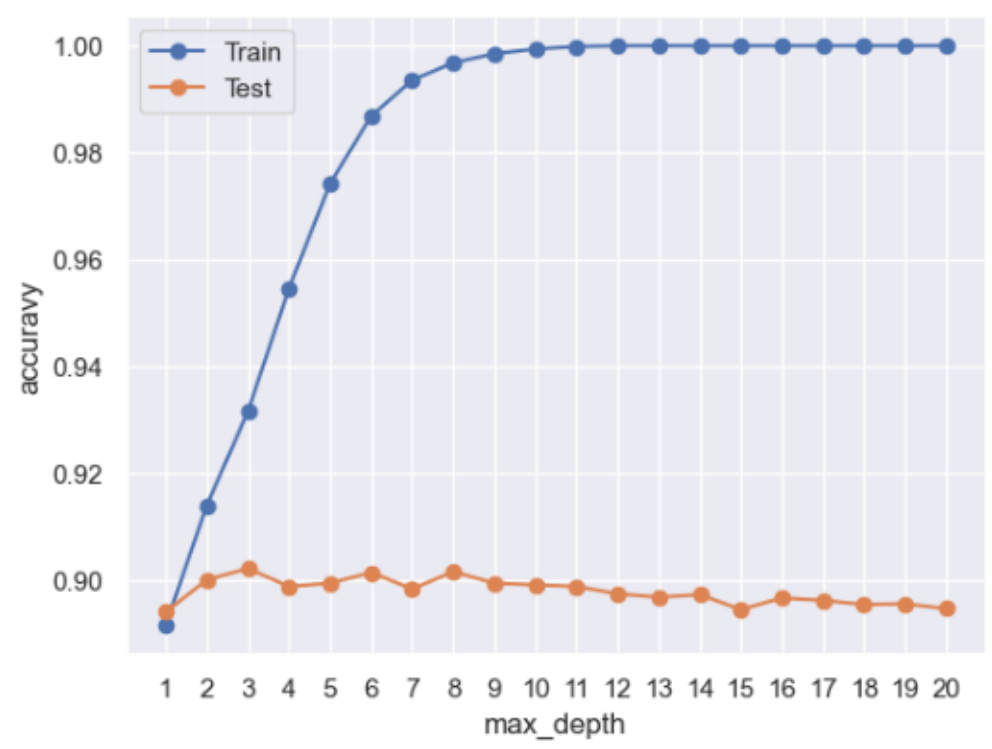

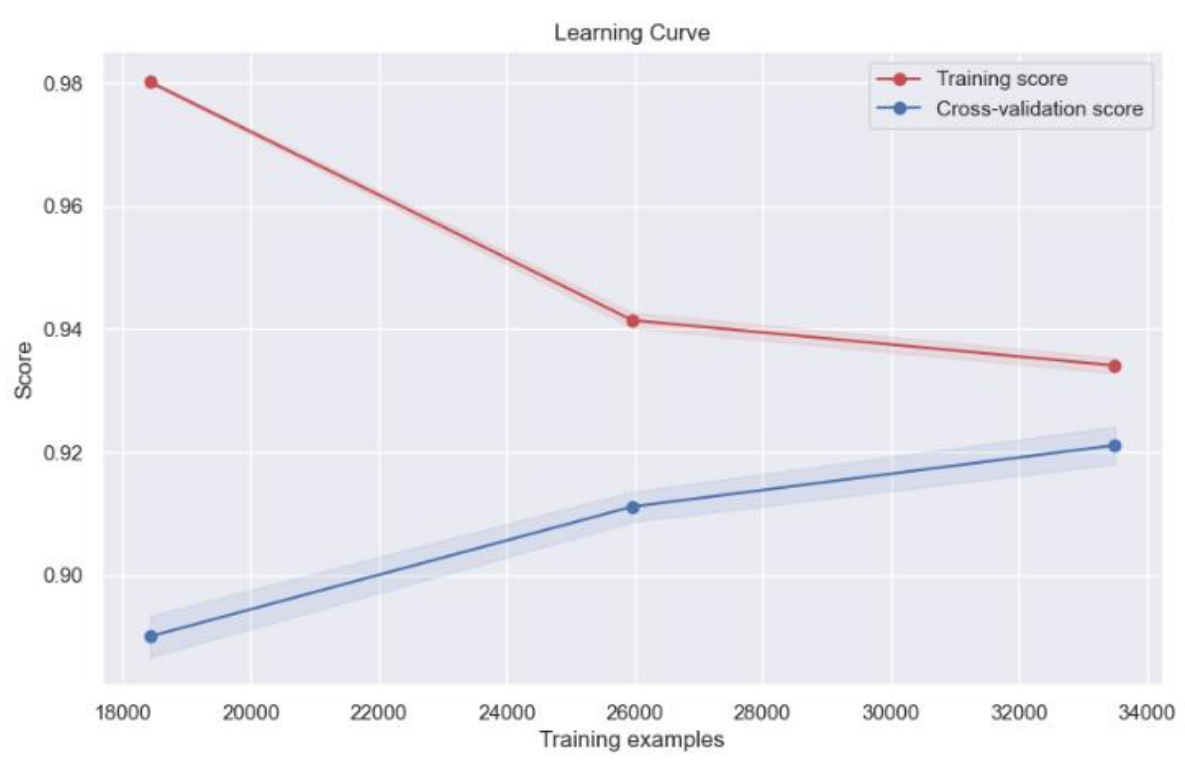

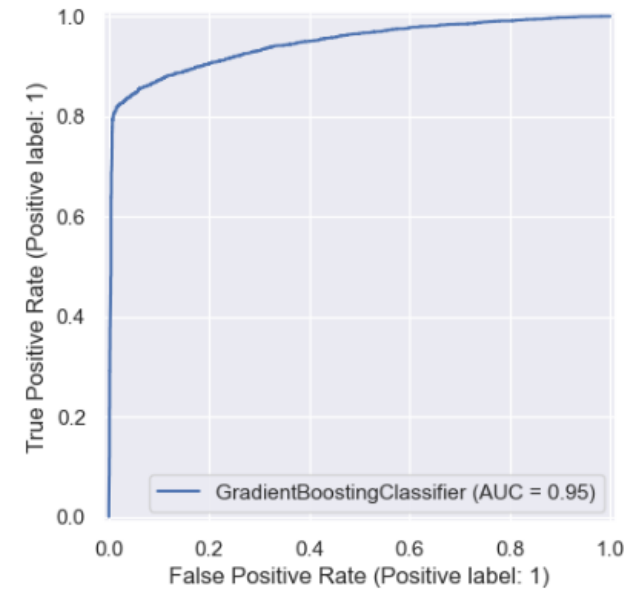

Figure 6. ML model based on PubChem bioassay AID2735, focusing on CHOP inhibitors

a) Tracing the deviation between train and test accuracy

b) Learning curve

c) AUC

The classification report for the XGBC demonstrates excellent overall performance with a high global accuracy of 90% across 6,500 samples (Table 6). Notably, the dataset is perfectly balanced, with equal support for both the Active (target 1) and Inactive (target 0) classes (3,250 samples each). The model demonstrates a clear priority for the Active (target 1) class, achieving an outstanding Recall of 98%. This means the model is exceptionally sensitive, correctly identifying 98 out of every 100 actual active compounds-a critical feature for minimising FNs in screening and drug discovery. This high Recall is paired with a good Precision of 85%, resulting in an F1-score of 0.91. Conversely, the Inactive (target 0) class shows a significant Precision of 97%, indicating high reliability when the model predicts a compound is inactive. However, its Recall is lower at 83%, suggesting that 17% of actual inactive compounds are incorrectly classified as active (FP). The convergence of the macro and weighted average F1-scores to 0.90 confirms the model's high utility and the overall stability of its predictive performance.

Table 6. Classification report for XGBC based on PubChem bioassay AID2735, focusing on CHOP inhibitors

| | precision | recall | F1-score | support |
|---|---|---|---|---|
| Active (target 1) | 0.85 | 0.98 | 0.91 | 3250 |
| Inactive (target 0) | 0.97 | 0.83 | 0.89 | 3250 |
| | | | | |
| accuracy | | | 0.90 | 6500 |
| macro avg | 0.91 | 0.90 | 0.90 | 6500 |
| weighted avg | 0.91 | 0.90 | 0.90 | 6500 |

The robustness analysis, benchmarked against an excellent Baseline Accuracy of 0.9017, reveals that the Gradient Boosting Classifier's high performance is highly contingent on data quality. The model demonstrates extreme sensitivity to Gaussian Noise, with accuracy plummeting by nearly 29% to 0.6432 at the lowest noise level (0.1) and stabilising barely above random chance at 0.5438 as noise increases. Conversely, the model shows moderate, though still declining, resilience when handling missing data via imputation, maintaining 0.8575 accuracy at a 10% missing rate before decreasing to 0.7326 at a 40% rate. This disparity indicates the model is vulnerable to minor feature perturbations but the imputation strategy is somewhat effective. Overall, the Average Robustness Accuracy is 0.6303, a significant 27% drop from the baseline, confirming that the model is not robust and requires rigorous data preparation to ensure predictive reliability in real-world applications.

The LIME analysis for the GBC provides clear evidence of the model's reliance on specific features for accurate classification, particularly for a successful True Negative prediction (Instance 3, True Label 0, Predicted Label 0). The model's decision to classify the compound as Inactive was overwhelmingly driven by the feature -1.08 < SID ≤ 0.24, which contributed a massive positive weight of +0.5525. This robust association confirms that the GBC has correctly learned to use this range of intermediate SID values as the primary indicator for inactivity in this local region. In contrast, the second feature, -0.26 < CID ≤ -0.17, offered only a negligible influence (+0.0053), demonstrating that the SID feature was the singular, dominant factor dictating the final classification outcome.

The comparison of the three ML models, based on CID_SID, MORGAN2, and RDKit-transformed SMILES features, clearly establishes the CID_SID ML model as the superior classifier, indicated by its commanding MCC of 0.8118, significantly higher than the RDKit-SMILES model (0.6474) and the MORGAN2 model (0.4961). This exceptional performance is rooted in the CID_SID model's confusion matrix, which shows the highest number of TP(2,683) and the lowest number of FN (75), yielding an outstanding precision of 97.28%. Conversely, the MORGAN2 model is the weakest, suffering from an extremely high number of FP (1,022), which severely cripples its precision and overall MCC. While the RDKit-transformed SMILES model achieves the highest number of TN (4,255) and the lowest number of FN (335), indicating a strength in identifying the negative class and capturing actual positives (recall), its high MCC and F1-score still place it second to the CID_SID model, which demonstrates the best overall balance of performance metrics.

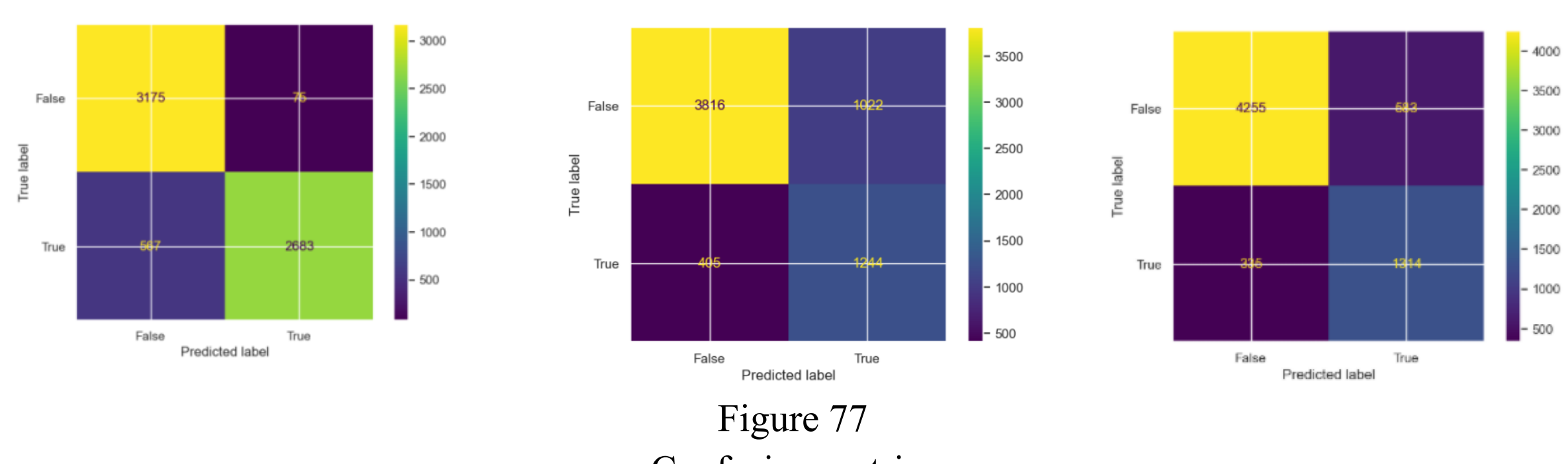

Figure 77
Confusion matrix:
a) CID_SID ML Model, b)ML based on MORGAN2 Features,
c)ML based on RDKit transformed SMILES

### 4. Results of the CID_SID ML model based on PubChem Bioassay AID588852, focusing on human M1 muscarinic receptor antagonists

The final dataset was constructed by merging the PubChem AID 588852 Bioassay (359,484 samples) with the PubChem AID 1996 Bioassay (57,859 samples), resulting in a dataset of 57,199 samples in total, 56,688 of which were inactive small biomolecules and 511 active. To continue reducing the inactive compound population, a systematic sampling approach was applied; every fourth sample from the population was extracted, creating a reduced set of 14,172 inactive samples. These were then integrated with all 4,560 active compounds to generate a final, consolidated dataset of 18,732 compounds for further use in the study. (60)

To ensure equal target representation and maximise model robustness, the test set was created by removing 10% of the active and inactive compounds, respectively. Specifically, 2,070 samples were randomly set aside from the inactive compounds after shuffling, and the same was done for the active compounds. These two classes were then integrated to form a balanced test set of 4,140 compounds, representing 20% of the initial consolidated dataset and featuring an equal proportion of targets. The remaining 14,592 compounds (18,732- 4,140 test) were designated for the training set. This training set was further processed by applying both scaling and balancing techniques (as illustrated in Figure 1). The imbalance in the training data was addressed using a RandomOverSapmpler, which effectively increased the number of compounds from 16,389 to 18,602 compounds, consequently ensuring the model was trained on a fully balanced dataset.

The metrics of the ML models listed in the Methodology section were compared, revealing that the XGBC was the most effective estimator, achieving the highest test set metrics: 82.7% Accuracy and 82.7% ROC, followed by GBC with 82.6% Accuracy and 82.6% ROC (Table 7).

Table 7. Metrics across the ML models based on PubChem Bioassay AID588852, focusing on human M1 muscarinic receptor antagonists

| Algorithm | Accuracy | Precision | Recall | F1 | ROC |
|---|---|---|---|---|---|
| XGBoost | 0.827 | 0.874 | 0.765 | 0.816 | 0.827 |
| GradientBoost | 0.826 | 0.874 | 0.761 | 0.814 | 0.826 |
| RandomForest | 0.799 | 0.919 | 0.656 | 0.765 | 0.799 |
| K-nearest | 0.797 | 0.841 | 0.733 | 0.783 | 0.797 |
| SVM | 0.795 | 0.858 | 0.708 | 0.776 | 0.795 |
| Decision | 0.780 | 0.892 | 0.638 | 0.744 | 0.780 |

Further analysis using five-fold cross-validation confirmed XGBC's robustness, yielding an average Accuracy of 86.4%±0.45, Precision of 58.01%±1.14, Recall of 73.57%±0.73, F1-score of 64.87%±0.94 and ROC AUC of 87.78%±0.47. The plot, Figure 8.a, illustrating the deviation between training and test accuracy for XGBC indicates a clear overfitting trend driven by increasing model complexity, specifically violating the established 5% threshold for the difference between training and test accuracy. The configuration with a complexity level of max_depth = 5 is superior, yielding a strong training Accuracy of 87.0% and a test accuracy of 83.2%, resulting in a generalisation gap of only 3.8%. This gap is well within the acceptable limit, indicating a balanced and robust model. However, increasing the complexity to max_depth=6 leads to the training accuracy rising to 89.0% while the test Accuracy paradoxically drops to 82.7%. This widening generalisation gap of 6.3% exceeds the 5% threshold, signalling significant overfitting where the model has begun to memorise training noise at the expense of its ability to generalise to new data. Therefore, the max_depth=5 configuration represents the optimal trade-off, balancing high predictive performance with strong generalisation. **The final XGBC, hyperparameter-tuned and checked for overfitting, achieved an Accuracy of 83.6%, a Precision of 88.1%, a Recall of 77.6%, an F1-score of 82.5% and an ROC of 83.6%.**

The analysis of the XGBC learning curve, which plots performance against the number of training examples (Figure 8.b), demonstrates a clear transition from high variance to a well-generalised state. Initially, with a small number of training examples (up to ~6,000), the model severely overfits the data, indicated by a perfect Training Score (1.0) and a very low Cross-validation Score (0.50). This wide gap shows the model is memorising noise rather than learning general patterns. As the training data volume approaches ~20,000 examples, the two curves rapidly converge: the Training Score gently decreases, while the Cross-validation Score sharply rises. At the maximum data point, both scores have largely plateaued and converged around 0.83. This convergence signifies that the model has reached the limit of its predictive capability given its current complexity and feature set, indicating that adding more training data alone is unlikely to yield significant further improvements in generalisation performance.

The plot (Figure 8.c), representing the ROC curve for the XGBC, demonstrates excellent discriminatory power as summarised by the AUC value of 0.89. The AUC metric, which ranges from 0.5 (random chance) to 1.0 (perfect classification), indicates that the model has an 89% chance of correctly ranking a randomly chosen positive sample higher than a randomly chosen negative sample. The curve's characteristic shape, rising steeply and hugging the top-left corner, visually confirms this high level of performance, showing that the model can achieve a high True Positive Rate while maintaining a low False Positive Rate across various classification thresholds. This AUC value is widely regarded as reflecting a very good or excellent classification model, confirming its reliability in distinguishing between the two target classes.

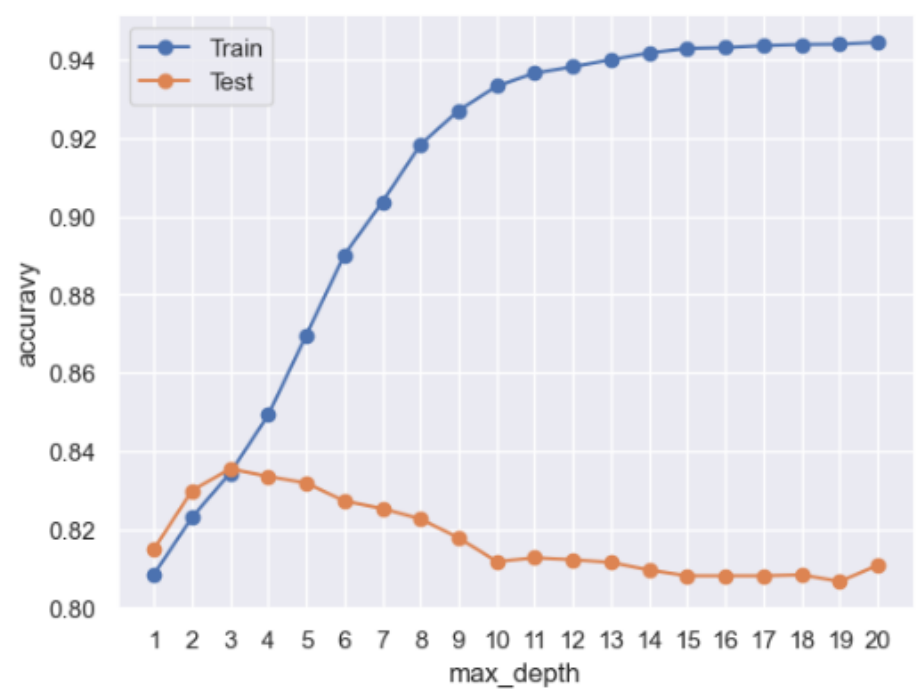

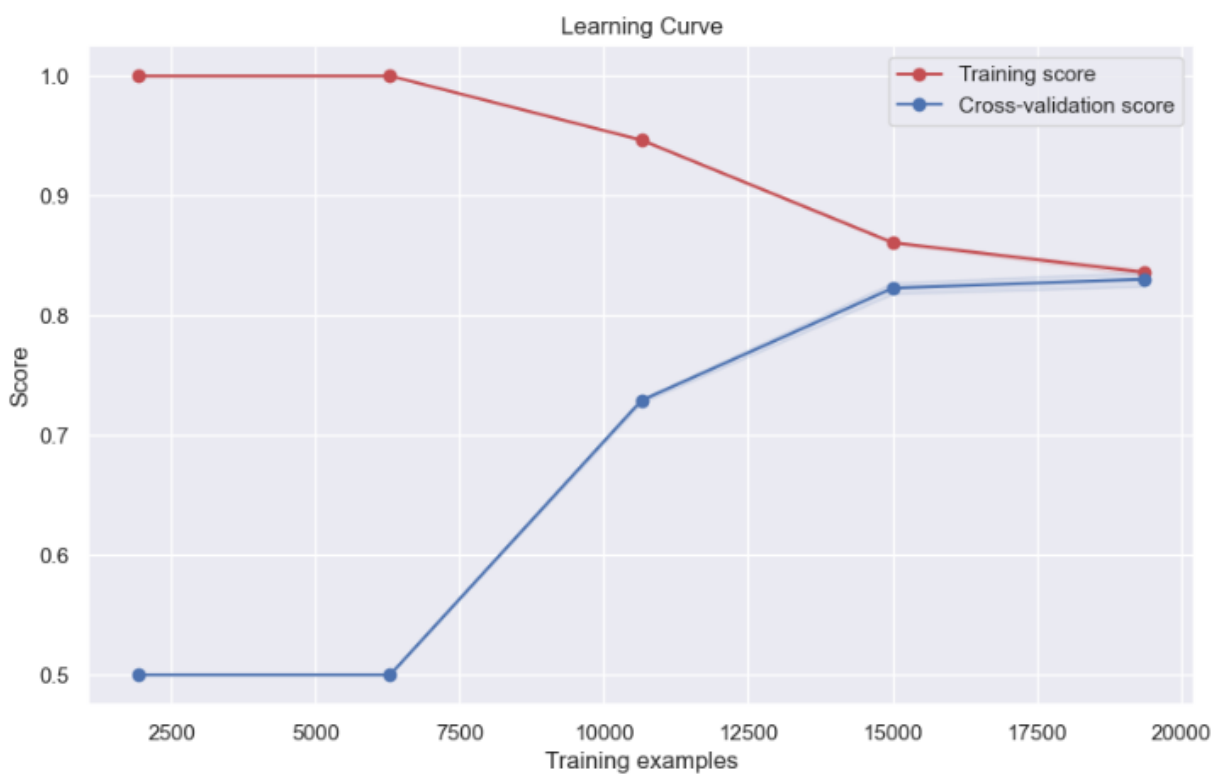

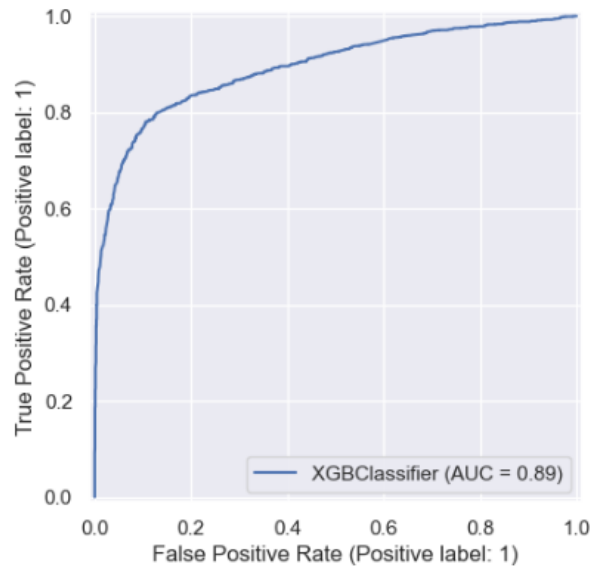

Figure 8. ML based on PubChem Bioassay AID588852, focusing on human M1 muscarinic receptor antagonists

a) Tracing the deviation between train and test accuracy
b) Learning curve
c) AUC

The provided classification report (Table 8) for the XGBC demonstrates strong overall performance, with a high global accuracy of 84%. With a high Recall of 89% for the Active (target 1) class, the model demonstrates significant proficiency, correctly identifying nearly nine out of ten true active compounds. The Precision for the Active class is 80%, resulting in an excellent F1-score of 0.84, which is particularly important for drug discovery, where minimising false negatives (missing an active compound) is critical. Conversely, the Inactive (target 0) class shows a lower Recall of 62%, suggesting the model is less effective at correctly identifying inactive compounds; however, it maintains a decent Precision of 84% when it does label a compound as inactive. This imbalance in recall is common in datasets with unequal class supports (2,750 Active vs. 2,070 Inactive, though not severely imbalanced) and further suggests that the model is slightly biased towards predicting the Active class, sacrificing some specificity for higher sensitivity. Overall, the macro and weighted average F1-scores of 0.83 confirm the model's high utility and effectiveness for the prediction task.

Table 8. Classification report for XGBC based on PubChem Bioassay AID588852, focusing on human M1 muscarinic receptor antagonists

| | precision | recall | F1-score | support |
|---|---|---|---|---|
| Active (target 1) | 0.80 | 0.89 | 0.84 | 2750 |
| Inactive (target 0) | 0.84 | 0.62 | 0.71 | 2070 |
| | | | | |
| accuracy | | | 0.84 | 4140 |
| macro avg | 0.84 | 0.84 | 0.83 | 4140 |
| Weighted avg | 0.84 | 0.84 | 0.83 | 4140 |

The robustness analysis of the XGBC reveals that its performance, starting from a high baseline accuracy of 0.8355 on clean data, degrades significantly under adversarial conditions. The model shows a distinct sensitivity to Gaussian Noise, with accuracy falling sharply from 0.6872 at a noise level of 0.1 to 0.5510 at a level of 0.6, indicating poor tolerance for random data perturbations. However, the model exhibits better, though still declining, resilience to missing values handled by imputation, with accuracy starting at 0.7932 when 10% of data is missing, and stabilising at 0.6804 with a 40% missing rate. Overall, the Average Robustness Accuracy is 0.6554, representing a substantial drop of nearly 18% compared to the baseline, which confirms that while the XGBC performs exceptionally well on clean data, its robustness is moderate to poor, requiring cautious deployment in real-world scenarios where data quality is uncertain.

The LIME analysis provides critical insight into the XGBoost Classifier's local decision-making by contrasting a successful prediction (Instance 0) with a prediction error (Instance 1), both having the same True Label of Inactive (0). For Instance 1, the model incorrectly predicted Active (1), a False Positive error driven by the feature SID ≤ -0.87, which contributed a strong positive weight of +0.485. This indicates a specific region in the feature space where the model's learned association is misleading. Conversely, for Instance 0, the model correctly predicted Inactive (0), a True Negative success largely attributable to the feature 0.25 < SID ≤ 0.37, which provided a strong negative weight of - 0.421. This comparison clearly shows that the model heavily relies on specific thresholds of the SID feature to make local classification decisions; while intermediate SID values are correctly associated with inactivity, highly negative SID values lead to a substantial prediction error.

The comparison of the three binary classification models based on their MCC, a robust metric for balanced performance, reveals that the ML model based on RDKit transformed SMILES is the most effective, with the highest MCC of 0.6861. This model achieves the optimal balance across the confusion matrix, evidenced by the highest number of (2579) and the lowest count of FN (189) (Figure 9) . The CID_SID ML model is highly competitive with an MCC of 0.6758. With 1607 TP, this model achieves the highest detection rate for the positive class; however, its 463 FN results in a lower  Recall and a higher proportion of missed positives than observed in the RDKit model. The ML model built on MORGAN2 features is the weakest performer, yielding an MCC of just 0.5606 due to a high rate of FP=396, which severely compromises its overall classification balance and utility.

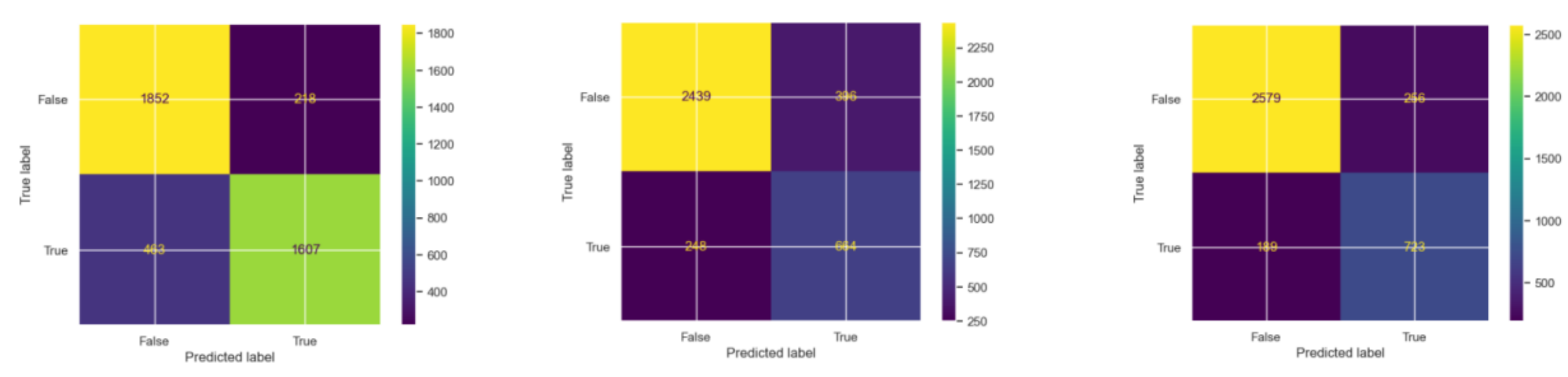

Figure 9.
Confusion matrix:
a) CID_SID ML Model, b)ML based on MORGAN2 Features,
c)ML based on RDKit transformed SMILES

The  ML strong and consistent performance across the four bioassays, achieving an average Accuracy of 83.52% (±5.56%) and an excellent ROC AUC of 89.6% (± 4.91%). The most notable characteristic is the model's prioritisation of Precision (87.9% ± 5.04%) over Recall (75.6% ±8.31%), indicating a high confidence in its positive predictions (low false positive rate) but suggesting it conservatively misses approximately 23% of true positive cases (a moderate false negative rate). Overall, the balanced performance is reflected in the solid F1-Score of 81.9% (±6.76%), confirming the model's suitability as a stable and reliable screening tool that generalizes well across the tested biological systems (Table 9 and Figure 10). The MCC, Table 10 and Figure 11a, reveals significant performance variability among the three machine learning (ML) descriptor methods, CID_SID, MORGAN2 features, and RDKit-transformed SMILES, across the four biological targets (D3, Rab9, CHOP, M1). The CID_SID ML model (blue) delivers the highest overall MCC score of 0.8178 for the CHOP inhibitors and also leads in performance for D3 dopamine receptor antagonists (0.7019). In contrast, the RDKit-transformed SMILES ML model (green) proves optimal for the remaining two targets, achieving the best scores for Rab9 promoters (0.6392) and M1 Muscarinic Receptor Antagonists (0.6862). The ML model based on MORGAN2 features (orange) consistently exhibits the weakest performance, recording the lowest or near-lowest MCC for all four targets, underscoring its general inefficiency in this context. In summary, no single descriptor method is universally superior; the blue method excels for CHOP and D3, while the green method is optimal for M1 and Rab9, demonstrating that performance is highly dependent on the specific biological target.

Table 9. Mean and standard deviation of ML metrics across the case studies based on CIDs and SIDs data

| | Accuracy | Precision | Recall | F1-score | ROC |
|---|---|---|---|---|---|
| D3 | 85.6 | 89.4 | 80.8 | 84.9 | 85.6 |
| Rab9 | 74.8 | 83.7 | 61.6 | 71.0 | 74.8 |
| CHOP | 90.1 | 97.3 | 82.6 | 89.3 | 90.1 |
| M1 | 83.6 | 88.1 | 77.6 | 82.5 | 83.6 |
| Mean | 83.52±5.56 | 89.62 ±4.91 | 75.65 ±8.31 | 81.93 ±6.76 | 83.53 ±5.56 |

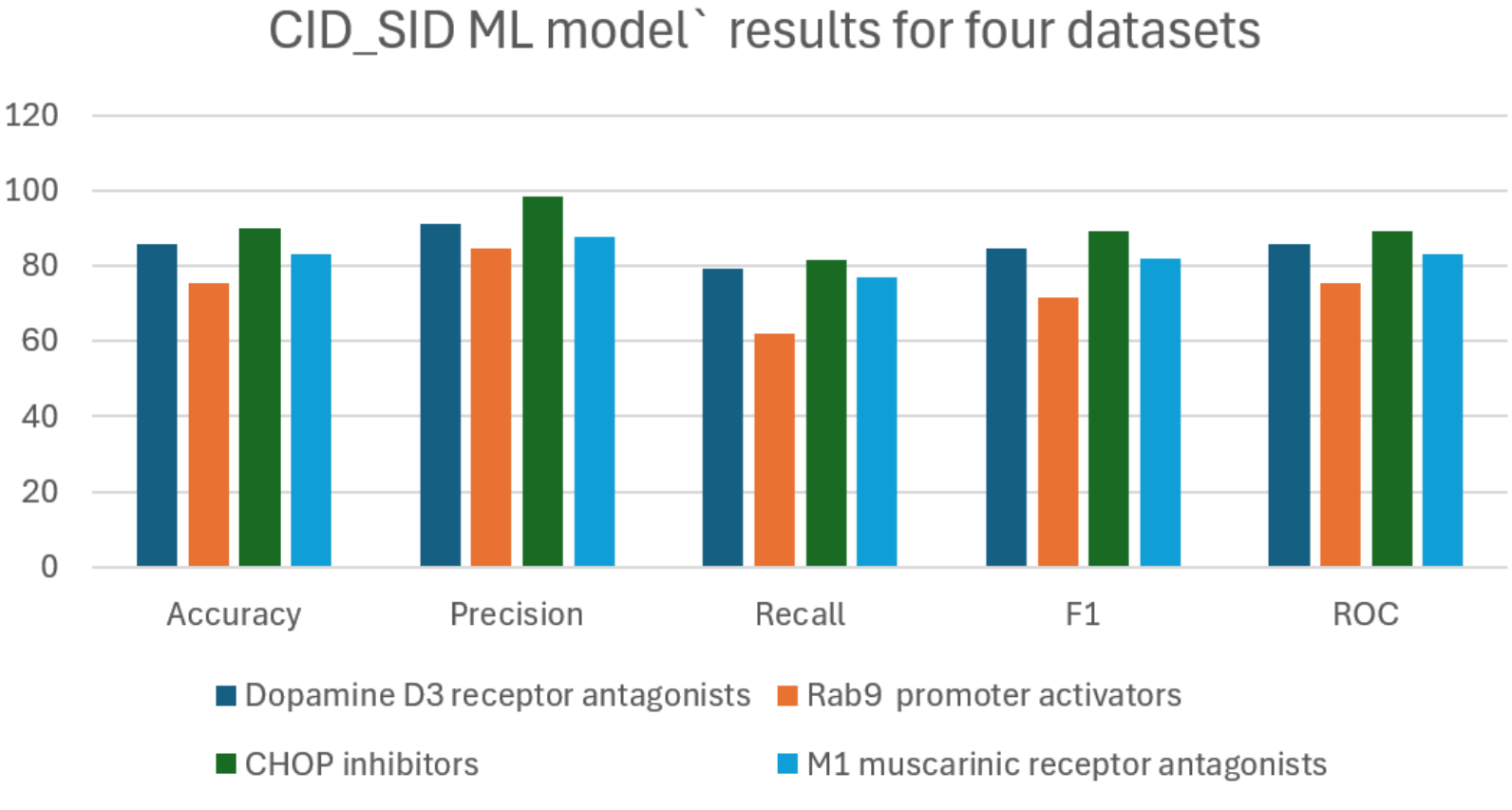

Figure 10.
ML metrics across all four CID_SID ML models

Table 10. MCC and time needed for running of the ML models

| | MCC | | | Time | | |
|---|---|---|---|---|---|---|
| | CID_SID | MORGAN2 | RDKit Smls | CID_SID | MORGAN2 | RDKit Smls |
| D3 | 0.7159 | 0.5833 | 0.6526 | 2.2285 | 18.4863 | 28.4926 |
| Rab9 | 0.5143 | 0.6377 | 0.6392 | 3.4846 | 140.4327 | 114.1453 |
| CHOP | 0.8118 | 0.4961 | 0.6474 | 5.1553 | 249.1418 | 277.5818 |
| M1 | 0.6758 | 0.5606 | 0.6862 | 2.3632 | 15.8985 | 23.7622 |

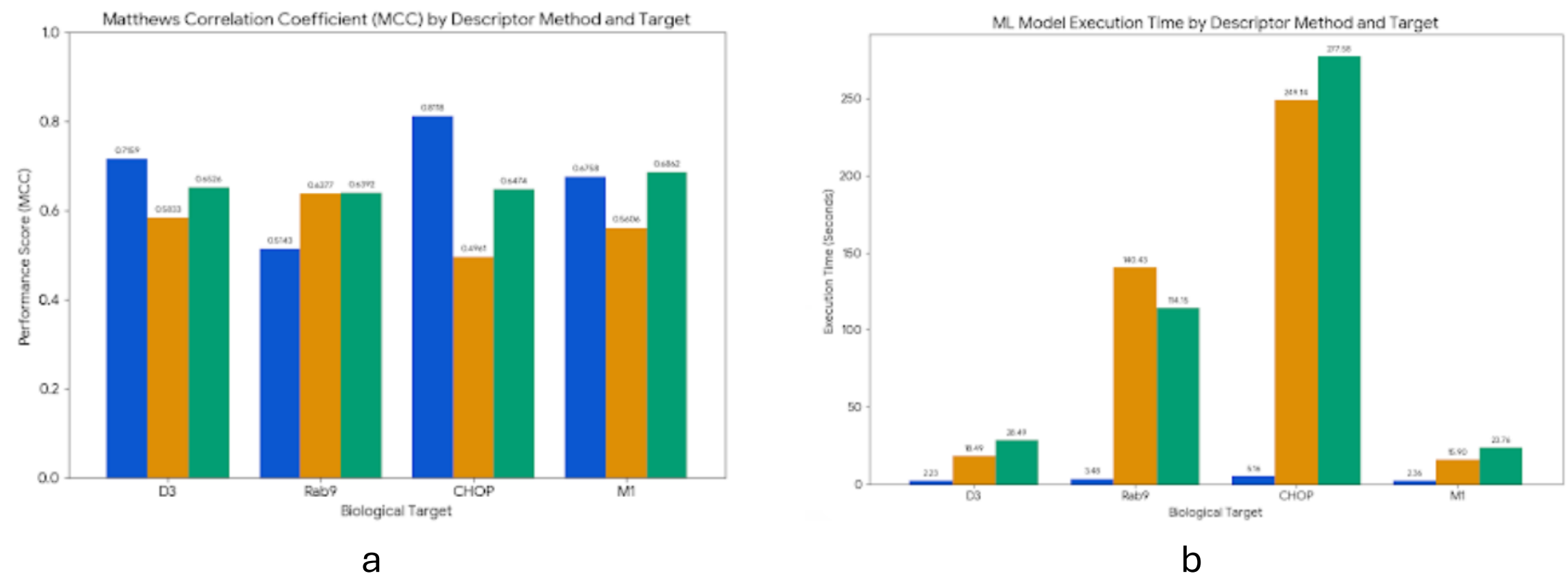

Figure 11.
Comparison between CID_SID ML model, ML model based on MORGAN2 features and NL model based on RDKit-transformed SMILES. a) MCC, b) Time for running the model from scaling to evaluation. Colour code: BLUE – CID_SID ML model; ORANGE -MORGAN2 ML model; GREEN -RDKit-transformed SMILES ML model

The execution time analysis, Table 10 and Figure 11.b, reveals a stark contrast in efficiency among the three ML models, where the CID_SID model (blue) is overwhelmingly superior, consistently registering the lowest execution times across all four biological targets (D3, Rab9, CHOP, M1), all under 5 seconds. For instance, it executes the D3 and M1 models in just 2.23s and 2.36s, respectively. In comparison, the ML model based on MORGAN2 features (orange) and the RDKit-transformed SMILES model (green) are significantly slower, with the latter showing the highest inefficiency by taking 277.58 seconds for the CHOP inhibitors target and also recording the slowest times for dopamine D3 receptor antagonists and M1 muscarinic receptor antagonists targets. While the RDKit-transformed SMILES model was previously shown to have superior predictive performance (MCC) for Rab9 and M1, its high computational cost, particularly the severe time penalty associated with CHOP, makes the CID_SID model the clear choice for applications prioritising rapid execution

## Discussion

Although the state-of-the-art (SOTA) in ML for drug discovery and development, particularly when dealing with molecular data, is dominated by Deep Learning approaches, with Graph Neural Networks (GNNs) and Generative Models as the most influential, the CID_SID model was deliberately compared against traditional ML models based on molecular descriptors derived from MORGAN2 and RDKit-transformed SMILES representations. This choice was made because the primary purpose of this study is to develop an entity suitable for large-scale prediction within a framework designed to predict millions of functionalities simultaneously. Even if a GNN would outperform the CID_SID ML model in terms of predictive accuracy, GNNs are inherently computationally expensive and are therefore not suitable for the required high-throughput, industrial-scale application. The focus was thus placed on developing a robust and scalable ML architecture capable of efficiently handling massive datasets

As noted above, no single descriptor method is universally superior; however, the CID_SID ML model substantially outperformed MORGAN2 and RDKit-transformed SMILES-based models regarding the time required for computation, validating its suitability as a highly efficient component for a scalable, high-throughput prediction framework. This efficiency was a direct and anticipated consequence of the data source: the feature computation, which converts structural data into numerical inputs, had already been performed by PubChem. This circumvents the time-intensive feature engineering step inherent to models utilizing MORGAN2 or RDKit, conferring a significant computational and temporal advantage to the CID_SID ML model methodology.

In the study, the generalisation of the model was thoroughly investigated. However, as hyperparameter tuning was performed with only a limited variation of hyperparameter values and trials, an exhaustive search for optimal hyperparameter settings can be considered as an option for further enhancing the robustness of the CD_SID machine learning model.

## Conclusion

The methodology presented in the study revealed that the information encoded in the PubChem SIDs and CIDs is beneficial beyond their identification task. The ML model results demonstrate that the methodology offers a time- and cost-effective approach for the early stage of drug development, with potential downstream benefits extending to clinical trials. Once the PubChem SID and CID have been obtained by a researcher for their small biomolecule, these identifiers will be enough for new functionalities of the compound to be predicted. For a demonstration of the idea and the approach in this study, four use cases were explored whose ML models can be directly used by researchers in drug discovery. Furthermore, the methodology is expected to be applicable to any PubChem bioassay which possesses a significant number of records and well-defined targets useful for ML training and testing. The primary limitation of this study is the lack of experimental confirmation. Due to restricted access to laboratory facilities, the proposed methodology remains theoretical and requires validation by an external team with the necessary experimental capabilities.

## Scientific contribution

The CID_SID ML model was developed as a fundamental building block for a time- and cost-effective ML framework. This framework is tailored to predict the functionality of small biomolecules beyond their initial design. This methodology is unique because it leverages PubChem identifiers (CID and SID), which inherently constrain AI hallucination and demonstrate strong potential for scalability across large chemical datasets.

## Acknowledge

MLI thanks the UWL Vice-Chancellor's Scholarship Scheme for their generous support. We sincerely thank the team at the National Center for Biotechnology Information (NCBI) and the developers and curators of the PubChem database for providing their comprehensive and freely accessible data. The article is dedicated to Luben Ivanov

## Author Contributions

MLI, MN, NR, GM and KN conceptualized the project and designed the methodology. MLI, NR and GM processed the data. MLI and NR wrote the code. KN supervised the project. All authors were involved with the writing of the paper.

## Data and Code Availability Statement

The Python code is available on GitHub as Jupyter notebook files:
https://github.com/articlesmli/CID_SID_ML_model.git

The datasets generated during the study are available on Hugging Face:

- Dataset on the human dopamine D3 receptor antagonists (57)
- Dataset on Rab9 modulators (58)
- Dataset on CHOP inhibitors (59)
- Dataset on M1 muscarinic receptor antagonists (60)

The raw data used in the study provided by PubChem:

- Dataset PubChem focused on dopamine D3 receptor antagonist (27)
- Dataset AID 485297 focused on the promoter of the protein Rab9 activators (28)
- Dataset PubChem AID 2732 focused on CHOP inhibitors (29)
- Dataset PubChem AID 588852 focused on M1 muscarinic receptor antagonists (30)
- Dataset PubChem AID 1996 focused on aqueous solubility of small biomolecules (33)

## Conflicts of Interest

The authors declare no conflict of interest.